# Does External Pressure Explain Recent Results for Molecular Clouds?

by

George B. Field, Eric G. Blackman, and Eric R. Keto

**Abstract**

The recent paper by Heyer et al (2009) indicates that observations of size, linewidth and column density of interstellar clouds do not agree with simple virial equilibrium (VE) as a balance between gravitational and kinetic energies in the sense that the clouds either have too much kinetic energy or too little mass to be bound. This may be explained by violation of VE as suggested by Dobbs et al 2011, by observational underestimation of the masses as suggested by Heyer et al 2009, or by an external pressure acting as an additional confining force as suggested earlier by Heyer et al 2004. The data of Heyer et al. 2009 cannot be explained with a single value for the external pressure, but if different clouds in the sample have different external pressures in the range of $P_e/k$ = E4 to E7 cm$^{-3}$ K, then most of the clouds could be in pressure virial equilibrium (PVE). In this paper we discuss two consequences of the external pressure. First, we show that the observational data are consistent with the hypothesis (Chié ze 1987) that most clouds are at a critical mass for dynamical stability determined solely by the pressure. Above this mass a cloud is unstable to gravitational collapse or fragmentation. Second, we show that the external pressure modifies the well-known size-linewidth relationship first proposed by Larson (1981) so that the proportionality is no longer constant but depends on the external pressure.

*1. Introduction*

The recent analysis of molecular clouds (MCs) by Heyer et al (2009) (Heyer09) challenges the power–law relationships first proposed by Larson (1981), between the size and linewidth $\sigma \propto R^{p_1}$ with $p_1 = 0.38$, and between the size and density, $\rho \propto R^{p_2}$ with $p_2 = -1.1$. Heyer09's analysis also challenges the consequent of these two relationships that the clouds are in simple virial equilibrium (VE) with the internal kinetic energy equal to half the gravitational energy. Previously, these several relationships seemed to be confirmed by other observational studies. Solomon et al (1987) (Solomon87) observed 273 GMCs and found $\sigma \propto R^{0.5}$, while Heyer & Brunt (2004) found $\sigma \propto R^{0.6 \pm 0.07}$ in 27 GMCs. In these studies the constant of proportionality, $V_0 = \sigma/R^{P_1}$, was found to be the same everywhere. In contrast, the Heyer09 analysis of newer data from the Galactic Ring Survey (GRS) (Jackson et al. 2006) found that MCs do not appear to be in virial equilibrium and that $V_0$ varies in proportion to the square root of the column density $\Sigma$. They suggest that the earlier data which seemed to support Larson's hypotheses are also consistent with this revised conclusion, but the quality of the earlier data was insufficient to make this evident.

In deriving this conclusion, Heyer09 directly compared clouds that were observed both in the earlier S87 survey and the recent GRS. Both surveys were made with the same telescope, the 14m Five College Radio Astronomy Observatory (FCRAO), but with significant differences. The more recent GRS survey was made with a 4x4 multi-pixel array camera (SEQUOIA) rather than a single pixel detector. The improvement in efficiency allowed the GRS to sample the clouds at about the Nyquist spacing for the 47" beam. In contrast, the S87 survey was undersampled with 3' spacing and missed many high-density structures which tend to be small scale. Additionally, the GRS survey observed $^{13}$CO rather than the more common $^{12}$CO used in the Solomon87 survey. The rarer isotopologue allows measurement of higher column densities before the line saturates. Summarizing the complete discussion in Heyer09, the better sampling and the tracer with lower optical depth enable the GRS survey to extend the relationship between $\Sigma$ and the quantity $V_0$ to higher column densities ($\Sigma > 100\ M_\odot \text{pc}^{-2}$).

The newer GRS data plotted in figure 7 of Heyer09 show that the column density, $\Sigma$, is not constant from cloud to cloud; that the scaling coefficient $V_0$ scales with column density as $V_0 = \sigma / L^{1/2} \propto \Sigma^{1/2}$; and that the clouds lie above the line of simple VE in the diagram of $V_0$ versus $\Sigma$. The first point suggests that Larson's third observation of constant column density (Larson 1981) requires reinterpretation. The second point shows that the size-linewidth relation is not a simple power-law dependency, and the third point implies that the clouds cannot be bound only by their self-gravity . For convenience we include this data as our Figure 1, where as in Heyer 2009, the line of simple virial equilibrium (VE) is the solid line.

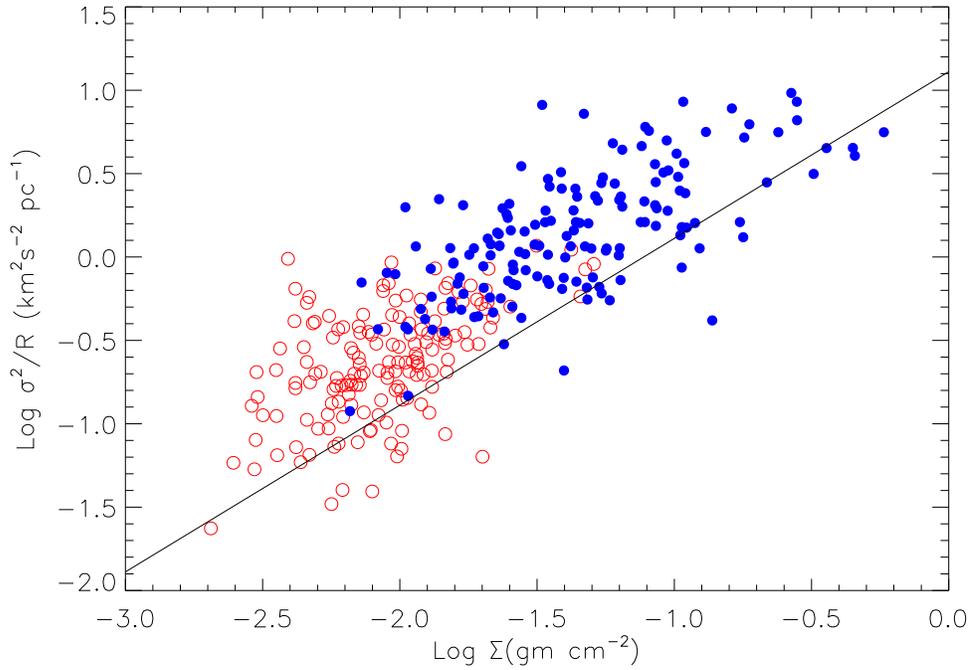

*Figure 1: Comparison of the Solomon87 and GRS surveys. The open-red and filled-blue symbols show the properties of the same clouds as derived from the Solomon87 and GRS surveys respectively. The comparison shows that better spatial sampling and use of a molecular tracer with lower optical depth detects clouds with higher column densities. The line represents simple virial equilibrium (VE). This figure is drawn from the data provided in Heyer09 and is similar to their Figure 7.*

Heyer09 noted that the trend in the data, parallel and to the left of the VE line, would be consistent with VE if the GRS survey systematically underestimated the cloud masses by a factor of 2 or 3 (0.3 or 0.5 dex). They suggested a number of reasons why this might be the case. An alternative explanation is offered by Dobbs et al (2011): that the data are correct, but in fact $\sigma^2$ is too large to satisfy VE. In other words, the clouds are unbound.

A third possibility is that the data are correct, but that external pressure $P_e$ provides an additional confining force not taken into account in the simple VE between gravitational and kinetic energies as suggested in a number of previous papers (Bonnor 1956, Keto & Myers 1987, Elmegreen 1989, Bertoldi & McKee 1992, Heyer & Brunt 2004, Lada et al 2008). In this paper, we explore the consequences of this external pressure.

In particular, the existence of an external pressure implies a critical mass above which a cloud cannot exist in stable equilibrium. We follow an earlier suggestion by Chie'ze (1987) that the masses of MCs tend to be equal to a critical mass $M_c$, defined as the largest stable mass for their internal kinetic energy and external pressure. If a cloud's mass is equal to its critical mass, then hydrostatic equilibrium (generally defined with support from internal kinetic as well as thermal energy) implies a critical column density that is now defined solely by the external pressure.

If the clouds in the interstellar medium tend toward this critical column density, a further consequence is that there is still a size - linewidth relation, but it is not a simple proportionality as originally suggested by Larson (1981). Rather the critical mass defines a critical radius with the result that the proportionality between linewidth and size depends on the external pressure, $\sigma^2/R \propto \sigma^2/R_c \propto P_e^{1/2}$.

## 2. The Critical Mass and the External Pressure

### 2.1 The virial theorem

The virial theorem that applies to an isolated self-gravitating isothermal spherical cloud immersed in a uniform external pressure, $P_e$, is (Spitzer 1978),

$$\frac{\ddot{I}}{2M} = 3\sigma^2 - \frac{\Gamma GM}{R} - \frac{4\pi P_e R^3}{M}, \qquad (1)$$

where $I = CMR^2$ is the moment of inertia, $C$ is a dimensionless constant whose value is not important in what follows, and $\Gamma$ is a form factor which equals 3/5 for a sphere of constant density and 0.73 for an isothermal sphere of critical mass (Elmegreen 1989), a state to be discussed further below. In equilibrium, $\ddot{I} = 0$. (Conventionally, $\ddot{I} < 0$ implies collapse and $\ddot{I} > 0$ implies expansion; however, see Va´zquez – Semadeni (1997) and Ballesteros – Paredes (1999) for critical discussions of this assumption.)

Equation 1 could contain another term representing magnetic energy; however, this term may not be important. First, a recent analysis (Crutcher et al. 2011) concludes that many fields are so weak as to be dynamically irrelevant. Second, observational analyses of survey data such as Solomon87 and Heyer09 do not consider magnetic energy. Third, in our analysis, neglect of this term may be justified *post facto*.

From equation 1, the critical mass can be derived in a number of ways, most simply by assuming that the cloud is a sphere of constant density ($\Gamma = 3/5$), and calculating the maximum equilibrium pressure for a given mass and kinetic. The result can also be interpreted as the maximum mass of a cloud in equilibrium for a given pressure and kinetic energy. From the condition that $dP_e/dR = 0$, we find the critical radius,

$$R_c = \frac{4GM}{15\sigma^2}, \qquad (2)$$

and the critical mass,

$$M_c = \left(\frac{3}{\pi}\right)^{1/2} \frac{15^{3/2}}{32} \frac{\sigma^4}{G^{3/2} P_e^{1/2}} = 1.77 \frac{\sigma^4}{G^{3/2} P_e^{1/2}} \qquad (3)$$

The relationship between the critical pressure for a given mass and the converse is illustrated in Figure 2, which shows the solutions of equation (1) as pressure versus radius. This type of figure is familiar from studies of the equilibrium of pressure - bounded (Bonnor-Ebert) spheres (e.g. Figure 1, Bonnor 1956). In the conventional interpretation, the equilibria on each curve to the right of the maximum pressure have $dP/dR < 0$, and so are stable because an increase in the external pressure causes a decrease in the volume and hence a counter-balancing increase in the internal density and pressure. Left of the pressure maxima, the volume is small enough that self-gravity dominates and the external pressure is irrelevant; a decrease in volume increases the inward gravitational force, and hence decreases the volume even further, leading to gravitational instability. The three curves in Figure 2 for three masses, 300, 600, and 900 $M_\odot$, show that for each mass there is a corresponding unique maximum pressure and the converse.

It is a question whether the clouds in the ISM can be in stable equilibrium in this conventional sense. Aside from the class of very small clouds known as starless cores (Ward-Thompson et al. 2007, Alves et al.

2001) whose internal energies are primarily thermal, clouds in the ISM, have internal energies dominated by supersonic motions which may not provide a conventional $dP/dR$ pressure-like resistance to contraction. In fact, the data of Heyer09 show that the clouds of the GRS survey conspicuously avoid the region of stable equilibrium (section 2.2), and this branch of the solution may therefore be irrelevant for these clouds.

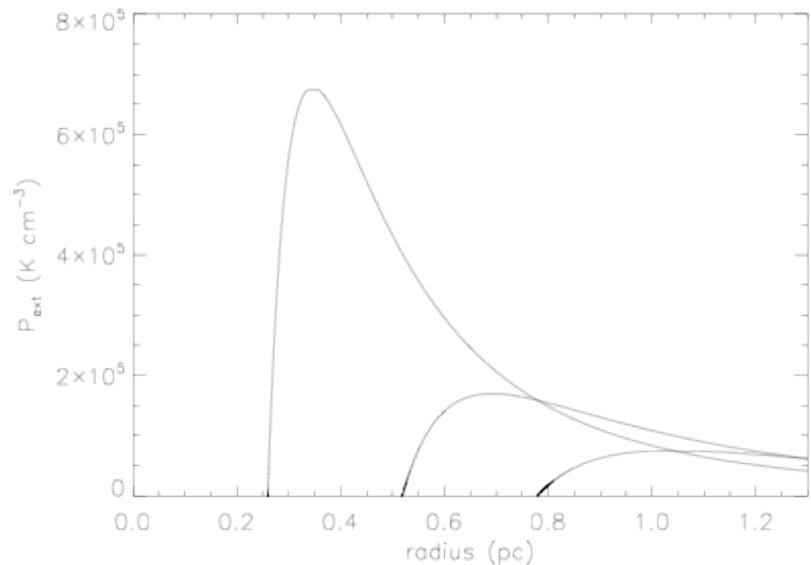

*Figure 2. Uniform density spheres in pressure-bounded virial equilibrium. The 3 curves show solutions of equation 1 for 3 different masses. To the right of the pressure maxima, $dP/dV < 0$, and the clouds are stable. To the left of the pressure maxima self-gravity dominates and the clouds must either collapse or fragment into smaller masses.*

The assumption of constant density is often used to analyze observations, but we know from observations, for example the distribution of mass with extinction (Lada et al 2009), that clouds have outwardly-decreasing internal density gradients. Williams et al (1995) suggest that the density inside clouds scales as $r^{-2}$. Because this is similar to the density gradient of a BE sphere (away from the center and edge), a better approximation to the critical radius and mass may be determined from the Lane-Emden (LE) equation for hydrostatic equilibrium. According to Bonnor (1956),

$$R_c = 0.49 \frac{GM}{\sigma^2}, \qquad (4)$$

and implies,

$$M_c = 1.2 \frac{\sigma^4}{G^{3/2} P_e^{1/2}}. \qquad (5)$$

(See Chi'eze (1987), Elmegreen (1989), and Holliman & McKee (1993) for other derivations.)

We combine equations (4) and (5) to obtain a critical column density,

$$\Sigma_c = M_c / \pi R_c^2 = \frac{M_c \sigma^4}{\pi (0.49)^2 GM_c^2} = 1.1 \left(\frac{P_e}{G}\right)^{1/2} = 5.1E-5 \left(\frac{P_e}{k}\right)^{1/2} \text{ g cm}^{-2} = 0.26 \left(\frac{P_e}{k}\right)^{1/2} M_\odot \text{ pc}^{-2}$$

(6)

and a critical value for the scaling coefficient

$$(V_{0,c})^2 = \frac{\sigma^2}{R_c} = \frac{1}{3}\left(\pi \Gamma G \Sigma_c + \frac{4P_e}{\Sigma_c}\right) = 6E-12 \left(\frac{P_e}{k}\right)^{1/2} \text{ cm s}^{-2} = 1.9E-3 \left(\frac{P_e}{k}\right)^{1/2} \text{ km}^2 \text{ s}^{-2} \text{ pc}^{-1} \qquad (7)$$

.

In terms of the two variables column density $\Sigma$ and the size-linewidth scaling coefficient $V_0$, the solutions of pressure-bounded virial equilibrium (PVE) for a single external pressure are V-shaped (asymptotic to linear with $\Sigma$ and $1/\Sigma$) as can be seen by rewriting equation (1) in the form,

$$V_0^2 = \frac{\sigma^2}{R} = \frac{1}{3}\left(\pi \Gamma G \Sigma + \frac{4P_e}{\Sigma}\right) \qquad (8)$$

**2.2 The observations**

In Figure 3 we show these solutions of pressure-bounded virial equilibrium along with the data from Heyer09. In computing the curves of PVE, we use $\Gamma = 0.73$ appropriate for clouds with a centrally concentrated internal density structure approximated by hydrostatic equilibrium rather than constant density, $\Gamma = 0.60$, although this makes a negligible difference on the solution curves for the range of densities in Figure 3. We omit the data points from Heyer09 in Figure 1 that correspond to the older Solomon87 survey. The 3 V-shaped curves show solutions for three different pressures. For the largest values of $\Sigma$ the clouds are dominated by self-gravity and their equilibrium curves are asymptotic to the solutions of simple VE (no external pressure) shown as the straight solid line.

The data are not consistent with PVE for any single value of the external pressure because the data do not lie along any of the V-shaped curves. Therefore, we should not expect any single estimate of the pressure of the inter-cloud medium to be generally applicable. Rather if pressure is important in the equilibria of clouds, then the individual clouds in the GRS survey must be in different environments with different external pressures.

If we suppose that the GRS clouds exist in a range of different pressure environments, the clouds could still lie anywhere in the region of equilibrium. In figure 3, this region is above and left of the line of SVE.

Referring back to our discussion in section 2.1, the absence of clouds in this region of the plot means that the clouds do not populate the stable branch of the solutions of equation 1 for any external pressure.

Rather, the clouds are clustered around the inflection points of the V-shaped curves of pressure equilibria and around a line defined by the critical column density and critical radius. We plot the line of points, $\sigma^2/R_c$ versus $\Sigma_c$, as a dashed red line in the log-log space of figure 3. The location of this line relating the critical density, radius, and mass depends on the internal density structure assumed for the cloud. For example, the line drawn in figure 3 assumes the density structure consistent with the Lane-Emden equation for hydrostatic equilibrium (section 2.1). If we assume a uniform internal density as is commonly done in analyzing observational data, the relationship between the radius, density, and mass is slightly different. In this case the line shifts to the right as shown by the difference between the asterisks (Lane-Emden) and diamonds (constant density) in figure 3.

The exact location of this line on the figure is less important than the observation that the clouds do not fill out the allowed space of equilibrium, the upper-left of the figure. Were we to adopt different approximations, for example a different geometry describing the clouds, the line and the data would shift yet again. Both the Lane-Emden and constant density approximations assume spherical geometry which does not describe the complex morphologies of the clouds. We use the spherical geometry because this approximation is almost universally employed in analyzing this type of observational data.

## 3. Implications

3.1 The lifetimes of clouds

Supposing that the clouds in the ISM tend toward their critical masses, why would this come about? In other words, what can we infer from this observation? In our previous paper we suggested that the properties of MCs are consistent with virial equilibrium including an external pressure, but that the equilibrium is unstable, leading to fragmentation and a continuous cascade of mass and energy to smaller scales.

In this interpretation, virial equilibrium implies a self-adjusting balance between the kinetic energy of internal motions and the potential energy of self-gravity and external pressure. We assume that the motions are driven primarily by gravitational forces so that an out-of-balance cloud may, for example, contract and generate higher velocities. Because the motions are chaotic, the contraction does not necessarily lead to collapse. This self-adjustment is possible even if the motions are partly, but not totally, driven by external forces such as bipolar outflows or supernovae.

Although a cloud may find its equilibrium, this does not imply that this state is long-lived. The cloud is continuously losing kinetic energy through dissipation and radiation. Since the critical mass depends on the internal kinetic energy, the point of criticality is continuously decreasing. Once the critical mass equals the actual mass, the cloud becomes unstable and either fragments into smaller masses or collapses. A smaller mass fragment in the same external pressure is further from the point of marginal stability (more stable) than its larger mass parent. However, the fragments are themselves losing energy through dissipation and soon find themselves at critical stability. This process occurs continuously on all scales in the MISM with the result that MCs on every scale are always at their critical masses.

This interpretation is different from the traditional interpretation of virial equilibrium as stable equilibrium and the lifetimes of the clouds as long-lived. In contrast, if the equilibrium is unstable as we suggest, then the relevant timescale is the dissipation time for each length scale, and a cloud exists as an individual entity only for this time. If the fragmentation cascade is in quasi-steady state, the dissipation time and cloud lifetimes are approximately the crossing-time at each length scale. In general, if clouds have short lifetimes we would not expect to see them in large numbers. However, because the clouds are continuously created from larger scales and destroyed to smaller scales, the ensemble of clouds in the fragmentation cascade can be relatively long-lived. At this point, we have not specified the origin of the clouds at the largest scale or the lifetime of the cascade.

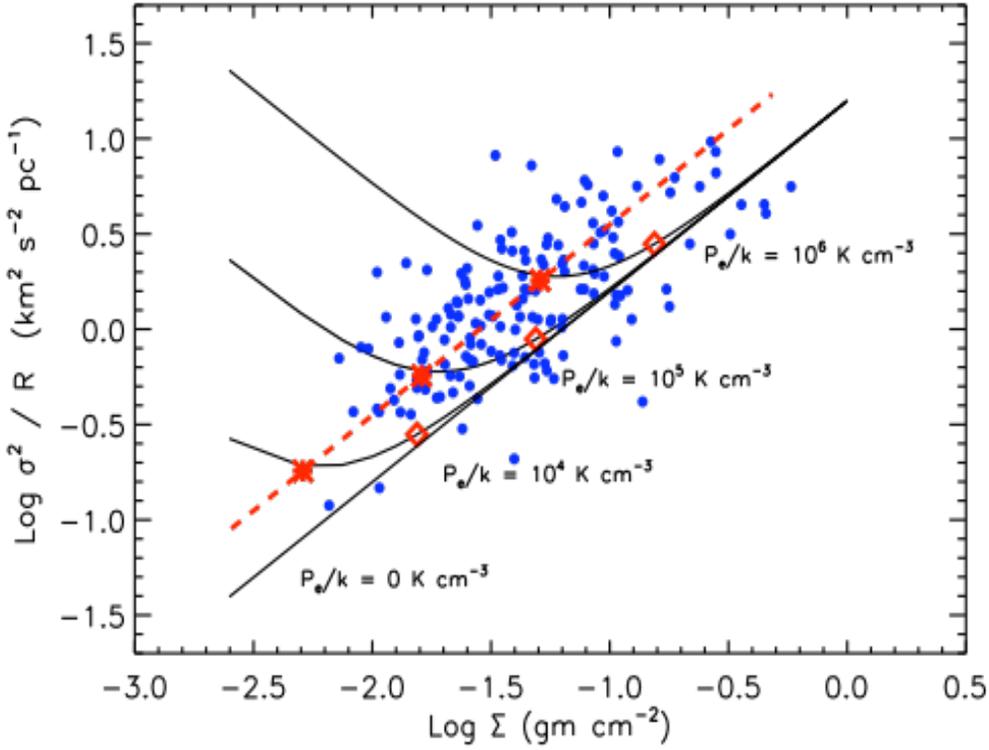

*Figure 3. Comparison of the GRS clouds with pressure-bounded virial equilibrium. The blue circles show the data from Heyer09. The three V-shaped curves are solutions of pressure-bounded virial equilibrium (PVE equation 1) for three different pressures as marked. The solid straight line shows the solution VE for no external pressure. The three asterisks show the location of clouds of critical mass for each of the three pressures with the critical mass and radius determined from the Lane-Emden (LE) equation (equations 6 and 7). The dashed straight line shows the location of clouds of critical mass corresponding to the full range of pressures. The diamonds correspond to clouds of critical mass determined with the approximation of constant density (CD) (equations 4 and 5). The difference between the asterisks and diamonds shows the difference between the CD and LE approximations in this analysis.*

3.2 The scaling relations

According to (7), Chie'ze's (1987) hypothesis that clouds should be at their critical mass and radius implies that the exponent $p_1$ in Larson's (1981) size-linewidth relation $V_0 = \sigma / R^{p_1}$, should be $½$. FBK showed that several scaling relations including those proposed by Larson (1981) can be derived from virial equilbrium and an assumption of one of the scaling relations, for example $p_1 = ½$. FBK assumed this value from observation. Chie'ze's hypothesis allows one to derive a model of fragmentation based solely on physical principles.

### *4.The Origin of the External Pressure*

The origin of the external pressure required to explain the observations as PVE is not yet certain. On theoretical grounds, Elmegreen (1989) estimated a typical pressure for the neutral ISM of $9E3$ K cm$^{-3}$. He argued that the confining pressure on a molecular cloud should be about $5E4$ K cm$^{-3}$ by combining the ISM pressure with the gravitation of the molecular cloud's HI halo. Field et al (2009) suggested that recoil pressure of the order of $E5$ would result from the release of H atoms from MCs by far UV radiation. Observations of individual regions suggest pressures of the order of $E5$ K cm$^{-3}$. For example, Bertoldi and McKee (1992) found pressures of $P_e/k = (0.5 - 2)E5$ K cm$^{-3}$ around molecular clouds in Ophiuchus, Lada et al (2008) found $7E4$ K cm$^{-3}$ in the Pipe Nebula, and Belloche et al (2011) found $P_e/k = 5E5$ K cm$^{-3}$ from observations of 60 nearby starless cores. These theoretical estimates and observations suggest intercloud pressures in the range required to confine most of the clouds, indicated by Figure 3 as between $E4$ and $E6$. We are not aware of other observations indicating higher pressures in the neutral ISM that might be necessary according to Figure 3.

## *5. Conclusions*

The data of Heyer et al (2009) challenge our understanding of the dynamics of molecular clouds. The clouds appear to systematically differ from the predictions of simple virial equilibrium based on the neglect of external pressure $P_e$ as a confining force. When corrected for that neglect, the clouds are in virial equilibrium if values of external pressure $P_e/k$ in the range $E4 - E6$ are assumed.

Pressure-bounded virial equilibrium by itself is not sufficient to explain the trend observed in the data. However, Chie'ze's (1987) additional constraint that cloud masses generally equal the critical value for dynamical instability allows a unique description that is consistent with the data. The clouds may be kept at the point of critical stability by dissipation and fragmentation to smaller scales. The observations of Heyer et al (2009) appear consistent with the FBK model of a gravitational fragmentation cascade provided that the masses of clouds have not been underestimated and that sources of external pressure can be identified.

*References*

Alves, J., Lada, C & Lada, E, 2001, Nature, 409, 159

Ballesteros-Paredes, J., Va'zquez-Semadeni, E., Scalo, J. 1999, ApJ, 515, 286

Belloche, A., Schuller, E., Parise, B., Hatchell, J., Jorgensen, J., Bontemps, S., Weisz, A., Menten, K., Muders, D, 2011, arXiv: 1101.0718

Bertoldi, F., McKee, C.F. 1992, ApJ, 395, 140

Bonnor, W. 1956, MNRAS, 116, 351

Chie'ze, J. P. 1987, A&A,171, 225

Crutcher, R., Wandelt, B., Heiles, C., Falgarone, E., Troland, T. 2010, ApJ, 725, 466

Dobbs, C., Burkert, A., Pringle, J. 2011, arXiv: 1101.3414

Elmegreen, B. 1989, ApJ, 338, 178

Field, G., Blackman, E., Keto, E. 2007, MNRAS, 385,181 (FBK)

Field, G., Blackman, E., Keto, E. 2009, arXiv: 0904.4077

Heyer, M., Brunt, C.M. 2004, ApJ, 615, L45

Heyer, M., Krawczyk, C., Duval, J., Jackson, J. 2009, ApJ, 699, 1092

Jackson, J. and 10 coauthors 2006, ApJS, 163, 145

Hoyle, F, 1953, ApJ, 118, 513

Keto, E., Myers, P. 1986, ApJ, 304, 466

McKee, C., Holliman, J. 1999, ApJ, 522, 313

Keto, E., Myers, P. 1996, ApJ, 304, 466

Lada, C., Muench, A., Rathborne, J., Alves, J., Lombardi, M. 2008, ApJ, 672, 410

Lada, C., Lombardi, M., and, Alves, J., 2009, ApJ, 703, 52

Larson, R. 1981, MNRAS, 194, 809

Solomon, P., Rivolo, A., Barrett,J., Yahil, A. 1987, ApJ, 319, 730

Spitzer, L., Jr. 1978, Physical Processes in the Interstellar Medium, Wiley – Interscience, New York

Va'zquez – Semadeni, E. 1997, astro-ph, 9701049

Ward-Thompson, D., Andre, P., Crutcher, R., Johnstone, D., Onishi, T., and Wilson, C., 2007, Protostars and Planets V, B. Reipurth, D. Jewitt, and K. Keil (eds.), University of Arizona Press, Tucson, 951, p.33-46,

Williams, J., Blitz, L., and Stark, A., 1995, ApJ, 451, 252